\documentclass[onecolumn,preprintnumbers,preprint,superscriptaddress,11pt,floatfix,nofootinbib]{revtex4-1}
\pdfoutput=1

\usepackage{amsmath}
\usepackage{graphics}
\usepackage{graphicx}
\usepackage{color}
\usepackage{subfig}
\usepackage[mathcal]{eucal}
\usepackage{cancel}

\begin{document}

\preprint{COLO-HEP-580, UCI-TR-2013-14, LA-UR-13-24696}

\title{Large $\theta_{13}$ in a SUSY $SU(5) \times T^{\prime}$ Model}

\author{Mu-Chun Chen}
\email[]{muchunc@uci.edu}
\affiliation{Department of Physics \& Astronomy, 
University of California, Irvine, CA 92697-4575, USA}
\author{Jinrui Huang}
\email[]{jinruih@lanl.gov}
\affiliation{Theoretical Division, T-2, MS B285,
Los Alamos National Laboratory, Los Alamos, NM 87545, USA}
\author{K.T. Mahanthappa}
\email[]{ktm@pizero.colorado.edu}
\affiliation{Department of Physics, University of Colorado at Boulder, Boulder, CO 80309-0390, USA}
\author{Alexander M. Wijangco}
\email[]{awijangc@uci.edu}
\affiliation{Department of Physics \& Astronomy, 
University of California, Irvine, CA 92697-4575, USA}

\date{\today}

\begin{abstract}
In the model based on SUSY $SU(5)$ combined with the $T^{\prime}$ family symmetry,  it is shown~\cite{Chen:2007afa} that fermion mass hierarchy and mixing angles can be naturally generated in both lepton and quark sectors compatible with various experimental measurements. But the predicted value for the non-zero lepton mixing angle of $\theta_{13} \simeq \theta_c/3\sqrt{2}$ contradicts the recent experimental results from Daya Bay and RENO. We propose to introduce one more singlet flavon field in the neutrino sector, and with its additional contribution, we are able to generate the large mixing angle $\theta_{13} \sim 8^{\circ}-10^{\circ}$. 
The analytical expressions of the mixing angles and neutrino masses with the additional flavon field are derived. 
Our numerical results show that a large region in the model parameter space is allowed for the normal hierarchy case, while a much smaller region is allowed for the inverted hierarchy case. In addition, the predicted value of the solar mixing angle is not affected by the additional singlet contribution. On the other hand, there exists a correlation between $\theta_{13}$ value and $\theta_{23} - \pi / 4$, as a result of the additional singlet contribution.      
We also make predictions for various other variables that can be potentially tested in the future neutrino experiments such as the Dirac CP phase, which is predicted to be in the range of $\delta \sim (195^{\circ} - 200^{\circ})$, and the neutrinoless double beta decay matrix element, $\left<m_{\beta\beta}\right>$. One numerical example with the minimal $\chi^2$ fit is presented, and the model predictions are consistent with all  experimental results.
\end{abstract}

\maketitle

\section{Introduction}

Quarks as well as leptons are known to mix, and the collective structure of these flavor oscillations has been a standing puzzle for decades. Since the discovery of neutrino oscillation, it is now well known that this phenomena could be described by a unitary matrix of at least rank three. Recent hints from T2K~\cite{Abe:2011sj}, Minos~\cite{Adamson:2011qu}, Double-Chooz~\cite{Abe:2011fz}, and discovery from Daya Bay~\cite{An:2012eh} and RENO~\cite{Ahn:2012nd} have now established a large value for the $\theta_{13}$ mixing angle in the neutrino sector. 
Existing global fits to a suite of neutrino oscillation experiments are summarized in the Table~\ref{tbl:exp_results} for both normal hierarchy (NH) and inverted hierarchy (IH) scenarios~\cite{Tortola:2012te} (see also, ~\cite{Fogli:2012ua, Morisi:2012fg,GonzalezGarcia:2012sz} and ~\cite{Machado:2011ar, Schwetz:2011zk, Schwetz:2011qt}). The experimental observation of a large lepton mixing angle $\theta_{13} \simeq 8^{\circ}-10^{\circ}$ is clearly incompatible with the so-called Tri-Bimaximal Mixing (TBM) pattern~\cite{Harrison:2002er},
\begin{equation}
U_{\mbox{\tiny{TBM}}} = \left(\begin{array}{ccc}
\sqrt{2/3} & \sqrt{1/3} & 0 \\
-\sqrt{1/6} & \sqrt{1/3} & -\sqrt{1/2} \\
-\sqrt{1/6} & \sqrt{1/3} & \sqrt{1/2}
\end{array}\right) \; , 
\end{equation}
which yields the following values for the mixing angles,
\begin{equation}
\sin^{2} \theta_{23}^{\mbox{\tiny TBM}} = 1/2 \;, \quad \tan^{2}\theta_{12}^{\mbox{\tiny TBM}} = 1/2 \;,  \quad \sin\theta_{13}^{\mbox{\tiny TBM}} = 0\ .\; 
\end{equation}

There have been many attempts to address the problem of fermion mass hierarchy and mixing.  A particularly popular approach has been through the introduction of a family symmetry, with which the number of free parameters in the Yukawa sector can be reduced. 
The symmetries proposed include the continuous ones, such as $SU(2)$~\cite{Ong:1978tq}, $SU(3)$~\cite{Kitano:2000xk} and $U(1)$~\cite{Chen:2006hn}, and the discrete ones~\cite{Altarelli:2010gt},
such as $A_4$~\cite{Ma:2001dn}, $S_4$~\cite{Hagedorn:2010th}, $T^{\prime}$~\cite{Frampton:1994rk}, $A_5$~\cite{Everett:2008et}. These discrete symmetries are particularly conducive to the TBM mixing pattern; in particular it has been shown that an additional $A_4$ symmetry can generate lepton mixing matrix naturally~\cite{Ma:2001dn}. However, there exists no simple model featuring an A4 symmetry which gives rise to quark mixing. As such a model would likely require additional mechanisms to be consistent with the quark sector~\cite{Ma:2006sk}, it is thus less compiling to combine A4 with a grand unified theory (GUT).

\begin{table}[t!]
\begin{tabular}{|c|c|c|} \hline\hline
parameter & $\text{best fit} \pm 1\sigma$ & $2\sigma$ \\ \hline
$\sin^2\theta_{12}$ & $0.320_{-0.017}^{+0.015}$ & 0.29-0.35 \\ \hline
$\sin^2\theta_{23}$ & $0.49_{-0.05}^{+0.08} (\text{NH});\; 0.53_{-0.07}^{+0.05} (\text{IH})$ & 0.41-0.62 (NH);\; 0.42-0.62 (IH) \\ \hline 
$\sin^2\theta_{13}$ & $0.026_{-0.004}^{+0.003} (\text{NH});\; 0.027_{-0.004}^{+0.003} (\text{IH})$ & 0.019-0.033 (NH);\; 0.020-0.034 (IH) \\ \hline
$\Delta m_{21}^2 (10^{-5}\text{eV}^{2})$ & $7.62\pm0.19$ & 7.27-8.01 \\ \hline
$\Delta m_{31}^2 (10^{-3}\text{eV}^{2})$ & $2.53_{-0.10}^{+0.08} (\text{NH});\; -(2.40_{-0.07}^{+0.10}) (\text{IH})$ & 2.34-2.69 (NH);\; -(2.25-2.59) (IH) \\ \hline
\end{tabular}
\caption{The current global fitting results on the neutrino mixing angles and mass splittings from~\cite{Tortola:2012te}. }   
\label{tbl:exp_results}
\end{table}

In light of the discovery of a large $\theta_{13}$, various models based on discrete family symmetries, 
such as $A_4$~\cite{Ma:2011yi,Cooper:2012wf, Chen:2012st, King:2013hj}, $S_4$~\cite{Hagedorn:2012ut}, $T^{\prime}$~\cite{Meroni:2012ty} and $A_5$~\cite{Cooper:2012bd} have been revisited lately, and more examples can be found in the review~\cite{King:2013eh}.
It has been pointed out~\cite{Chen:2012ha} that within the original $A_{4}$ models, due to the presence of the K\"ahler potential terms induced by the flavon vacuum expectation values (VEVs), there already exist ingredients that can generate a large value of $\theta_{13}$ compatible with current experimental observation. By going beyond the ``minimal'' setup with one triplet and one singlet flavon fields of the original $A_{4}$ models, it has also been shown~\cite{Chen:2012st} that the inclusion of additional $A_{4}$ singlets can also give sizable deviation from the TBM predictions, leading to phenomenologically viable predictions.

In the following, we consider the SUSY $SU(5)$ GUT model combined with the $T^{\prime}$ family symmetry proposed in~\cite{Chen:2007afa}. This model is free~\cite{Luhn:2008xh} of $T^{\prime}$ part of the discrete gauge anomalies~\cite{Araki:2008ek}  and can generate the fermion mass hierarchy and mixings in both lepton and quark sectors with nine real parameters, addressing the structure of flavor in a predictive way. In addition, leptogenesis can be realized by the geometrical CP phases~\cite{Chen:2007afa}. Unfortunately, while this model predicts a non-zero $\theta_{13}$ angle, the prediction is small compared to the experimental results. However, featuring unification, flavor structure, and a deterministic structure for CP violation, models of this class still merit consideration. The goal would then be to include a mechanism that raises the predicted value of $\theta_{13}$, which is the subject of this letter. Such efforts have been considered before; in a different $SU(5)\times T^{\prime}$ GUT model it has been demonstrated that one can preserve TBM in the neutrino sector by modifying the mixing in the charged lepton sector to generate an appropriate $\theta_{13}$ ~\cite{Meroni:2012ty}. In this letter, we take a different approach and introduce one more singlet scalar field coupling to the neutrino sector. Such an approach leads to deviations from TBM mixing, and the lepton mixing consistent with experimental values are obtained with relatively fewer parameters (namely ten, where all parameters are again real).

As the era of precision continues, it is the hope that the Dirac CP phase be measured and perhaps some insights on the conjugate nature of the neutrino can be gained through neutrinoless double beta decay searches. An excellent test of the model presented here can occur at both experiments, and to this end the predictions for both the Dirac CP phase, $\delta$, and the neutrinoless double beta decay matrix element, $\left<m_{\beta\beta}\right>$, are presented as well.

This paper is organized as follows. In Sec.~\ref{sec:model}, we present the model. The analytical and numerical results of the quark and charged lepton masses and quark mixings are discussed in the Sec.~\ref{sec:quark}. Both the analytical expressions and numerical results for the neutrino masses, lepton mixings and the experimental variables that may be accessible in the future experimental measurements are illustrated in the Sec.~\ref{sec:neutrino}. Section~\ref{sec:conclude} concludes the paper.

\section{The Model}
\label{sec:model}

We minimally expand the model ~\cite{Chen:2007afa} by adding a (non-trivial) singlet scalar field, $\eta^{\prime\prime} \sim 1^{\prime\prime}$. This additional singlet flavon field couples to the neutrino sector only, due to its charge under the discrete symmetries of the model.  The comprehensive particle content and their $SU(5)$, $T^{\prime}$ and $Z_{12} \times Z_{12}$ quantum numbers are summarized in Table.~(\ref{tbl:charge}).
\begin{table}[b!]
\begin{tabular}{|c|cccc|ccc|cccccc|cccc|}\hline
& $T_{3}$ & $T_{a}$ & $\overline{F}$ & $N$ & $H_{5}$ & $H_{\overline{5}}^{\prime}$ & $\Delta_{45}$ & $\phi$ & $\phi^{\prime}$ & $\psi$ & $\psi^{\prime}$ & $\zeta$ & $\zeta^{\prime} $ & $\xi$ & $\eta$ & $\eta^{\prime\prime}$ & $S$ \\ [0.3em] \hline\hline
SU(5) & 10 & 10 & $\overline{5}$ & 1 &  5 & $\overline{5}$ & 45 & 1 & 1 & 1 & 1& 1 & 1 & 1 & 1 & 1 & 1\\ \hline
$T^{\prime}$ & 1 & $2$ & 3 & 3 & 1 & 1 & $1^{\prime}$ & 3 & 3 & $2^{\prime}$ & $2$ & $1^{\prime\prime}$ & $1^{\prime}$ & 3 & 1 & $1^{\prime \prime}$ & 1 \\ [0.2em] \hline
$Z_{12}$ & $\omega^{5}$ & $\omega^{2}$ & $\omega^{5}$ & $\omega^{7}$ & $\omega^{2}$ & $\omega^{2}$ & $\omega^{5}$ & $\omega^{3}$ & $\omega^{2}$ & $\omega^{6}$ & $\omega^{9}$ & $\omega^{9}$ 
& $\omega^{3}$ & $\omega^{10}$ & $\omega^{10}$ & $\omega^{10}$ & $\omega^{10}$ \\ [0.2em] \hline
$Z_{12}^{\prime}$ & $\omega$ & $\omega^{4}$ & $\omega^{8}$ & $\omega^{5}$ & $\omega^{10}$ & $\omega^{10}$ & $\omega^{3}$ & $\omega^{3}$ & $\omega^{6}$ & $\omega^{7}$ & $\omega^{8}$ & $\omega^{2}$ & $\omega^{11}$ & 1 & $1$ & 1 & $\omega^{2}$
\\ \hline   
\end{tabular}
\vspace{-0.in}
\caption{
Field content of our model. The three generations of matter fields in $10$ and $\overline{5}$ of $SU(5)$ are the $T_{3}$, $T_{a}$ $(a=1,2)$ and $\overline{F}$ multiplets. 
The field $N$ contains three generations of right-handed neutrinos. Higgs fields that are needed to generate $SU(5)$ invariant Yukawa interactions are $H_{5}$, $H_{\overline{5}}^{\prime}$ and $\Delta_{45}$. The flavon fields $\phi$ through $\zeta^{\prime}$ 
are those that give rise to the charged fermion mass matrices, while $\xi$ through $S$ are the ones that generate neutrino masses.  The $Z_{12}$ charges are given in terms of the parameter $\omega = e^{i\pi/6}$.}  
\label{tbl:charge}
\end{table}
The charge assignments lead to the following Yukawa superpotential,
\begin{equation}
\label{eqn:superpot}
\mathcal{W}_{\mbox{\tiny Yuk}} =  \mathcal{W}_{TT} + \mathcal{W}_{TF} + \mathcal{W}_{\nu} \; ,
\end{equation}
where  
\begin{eqnarray}
\mathcal{W}_{TT} & = & y_{t} H_{5} T_{3} T_{3} + \frac{1}{\Lambda^{2}}  H_{5} \biggl[ y_{ts} T_{3} T_{a} \psi \zeta + y_{c} T_{a} T_{b} \phi^{2} \biggr] + \frac{1}{\Lambda^{3}} y_{u} H_{5} T_{a} T_{b} \phi^{\prime 3} \quad \label{eq:Ltt} ,\\ 
\mathcal{W}_{TF} & = &  \frac{1}{\Lambda^{2}} y_{b} H_{\overline{5}}^{\prime} \overline{F} T_{3} \phi \zeta + \frac{1}{\Lambda^{3}} \biggl[ y_{s} \Delta_{45} \overline{F} T_{a} \phi \psi \zeta^{\prime}  + y_{d} H_{\overline{5}^{\prime}} \overline{F} T_{a} \phi^{2} \psi^{\prime} \biggr]  \quad   \label{eq:Ltf} ,\\
\mathcal{W}_{\nu} & = & \lambda_{1} NNS +  \frac{1}{\Lambda^{3}} \biggl[ H_{5}  \overline{F} N \zeta \zeta^{\prime} \biggl( \lambda_{2} \xi + \lambda_{3} \eta   + \lambda_{4} \eta^{\prime\prime}
\biggr) \biggr] ,\quad\,
\label{eq:Lff}
\end{eqnarray}
with $\Lambda$ being the cutoff scale of the $T^{\prime}$ symmetry. The Yukawa couplings are real, which can be achieved through appropriate redefinition of the flavor fields~\cite{Chen:2007afa}. At the scale $\Lambda$, the $T^{\prime}$ symmetry is broken spontaneously and the flavon fields acquire vacuum expectation values (VEVs) along the following directions,
\begin{eqnarray}
\left<\xi\right> & = & \left(\begin{array}{c}
1 \\ 1 \\ 1
\end{array}\right)
\xi_{0} \Lambda , \,
\left< \phi^{\prime} \right> = \left(\begin{array}{c}
1 \\ 1 \\ 1
\end{array}\right) \phi_{0}^{\prime} \Lambda , \,  
\left< \phi \right> = \left( \begin{array}{c} 
0 \\ 0 \\ 1
\end{array}\right) \phi_{0} \Lambda , \, 
\left< \psi \right>  = \left( \begin{array}{c} 1 \\ 0 \end{array}\right)
\psi_{0} \Lambda , \,
\left< \psi^{\prime} \right>  =  \left(\begin{array}{c} 1 \\ 1 \end{array}\right) \psi_{0}^{\prime} \Lambda ,\\
\left< \zeta \right> & = & \zeta_{0} \Lambda \; , \qquad\quad \, \left< \zeta^{\prime} \right>  =  \zeta_{0}^{\prime} \Lambda ,\qquad \quad\;\,
\left< \eta \right> =  \eta_{0} \Lambda, \qquad \quad \;\, \left<S\right> =  s_{0} \Lambda, \qquad \quad \, \left<\eta^{\prime \prime}\right>  =  \eta_{0}^{\prime \prime} \Lambda\;.
\end{eqnarray}
We leave the study of the potential which gives rise to this alignment for future work. As such, we will not consider inputs from this sector in the counting of parameters needed to achieve the observed mixing, as these additional parameters in the flavon and Higgs sectors (such as $\tan\beta$) will lead to additional predictions such as the flavon and Higgs masses.

Upon the $T^{\prime}$ symmetry breaking, the operators in $\mathcal{W}_{TT}\;, \mathcal{W}_{TF}$, and $\mathcal{W}_{\nu}$ give rise to the masses of the up-type quarks, both the down-type quarks and charged leptons, and the neutrinos respectively.

\section{Quark and Charged Lepton Masses and Quark Mixings}
\label{sec:quark}
In terms of the $T^{\prime}$ and $SU(5)$ component fields, the above superpotential gives the following Yukawa interactions for the charged fermions in the weak charged current interaction eigenstates:
\begin{eqnarray}
\label{eq:L2}
-\mathcal{L}_{\mbox{\tiny Yuk}} &  \supset &  \overline{U}_{R, i} (M_{u})_{ij} Q_{L,j} + \overline{D}_{R,i} (M_{d})_{ij} Q_{L,j}  + \overline{E}_{R,i} (M_{e})_{ij} \ell_{L,j} + h.c. \; ,
\end{eqnarray}
where $Q_{L}$ denotes the quark doublets, $U_{R}$ and $D_{R}$ denotes the iso-singet up-type and down-type quarks, and $i$ and $j$ represent the generation indices. Similarly, $\ell_{L}$ and $E_{R}$ denote the iso-doublet and singlet charged leptons, respectively. 
The matrices $M_{u}$, $M_{d}$, and $M_{e}$, upon the breaking of $T^{\prime}$ and the electroweak symmetry, are given by~\cite{Chen:2007afa},
\begin{eqnarray}
M_{u} & = & \left( \begin{array}{ccc}
i y_{u} \phi^{\prime 3}_{0}  & 
(\frac{1-i}{2}) y_{u} \phi_{0}^{\prime 3} & 
0 \\
(\frac{1-i}{2})  y_{u} \phi_{0}^{\prime 3}  & 
y_{u} \phi_{0}^{\prime 3} + y_{c} \phi_{0}^{2} & 
y_{ts} \psi_{0} \zeta_{0} \\
0 & y_{ts} \psi_{0} \zeta_{0} & y_{t}
\end{array} \right) v_{u} \;,\\
M_{d}  & = & \left( \begin{array}{ccc}
0 & (1+i) y_{d} \phi_{0} \psi^{\prime}_{0} & 0 \\
-(1-i) y_{d} \phi_{0} \psi^{\prime}_{0} & y_{s} \psi_{0} \zeta^{\prime}_{0} \frac{v_{45}}{v_{d}}  & 0 \\
y_{d} \phi_{0} \psi^{\prime}_{0} & y_{d} \phi_{0} \psi^{\prime}_{0} & y_{b} \zeta_{0} 
\end{array}\right) 
v_{d} \phi_{0} \;  ,\\
M_{e} & = & \left( \begin{array}{ccc}
0 & -(1-i) y_{d} \phi_{0} \psi^{\prime}_{0} & y_{d} \phi_{0} \psi^{\prime}_{0} \\
(1+i) y_{d} \phi_{0} \psi^{\prime}_{0} & -3 y_{s} \psi_{0} \zeta^{\prime}_{0} \frac{v_{45}}{v_{d}} & y_{d} \phi_{0} \psi^{\prime}_{0} \\
0 & 0 & y_{b} \zeta_{0} 
\end{array}\right) 
v_{d} \phi_{0} .\; 
\end{eqnarray}
The $SU(5)$ symmetry in the model leads to the relation $M_{d} = M_{e}^{T}$, up to the $SU(5)$ CG's. Specifically, 
there is an additional factor of $-3$ in the (2,2) entry of $M_{e}$ due to the coupling to $\Delta_{45}$. In addition, the Georgi-Jarlskog (GJ) relations require $M_{e,d}$ to be non-diagonal, leading to corrections to the TBM pattern~\cite{Chen:2007afa}.  Note that the complex coefficients in the above mass matrices arise {\it entirely} from the CG coefficients of the $T^{\prime}$ group theory. More precisely, these complex CG coefficients appear in couplings that involve the doublet representations of $T^{\prime}$.

In total, the mass matrices in the charged fermion sector, $M_{u}$, $M_{d}$, and $M_{e}$, depend on seven independent parameters,
\begin{eqnarray}
\frac{M_{u}}{y_{t} v_{u}} & = & \left( \begin{array}{ccccc}
i g & ~~ &  \frac{1-i}{2}  g & ~~ & 0\\
\frac{1-i}{2} g & & g +  h  & & k\\
0 & & k & & 1
\end{array}\right)  , \\
\frac{M_{d}, \; M_{e}^{T}}{y_{b} v_{d} \phi_{0}\zeta_{0}}  & = &  \left( \begin{array}{ccccc}
0 & ~~ & (1+i) b & ~~ & 0\\
-(1-i) b & & (1,-3) c & & 0\\
b & &b & & 1
\end{array}\right) \; , 
\end{eqnarray}
where 
\begin{equation}
b \equiv \frac{y_{d}}{y_{b}} \frac{\phi_{0} \psi^{\prime}_{0}}{\zeta_{0}},  
\quad
c \equiv \frac{y_{s}}{y_{b}} \frac{\psi_{0}\zeta_{0}^{\prime}}{\zeta_{0}} \frac{v_{45}}{v_{d}}, 
\quad 
k \equiv \frac{y_{ts}}{y_{t}} \psi_{0}\zeta_{0}, \quad
h \equiv \frac{y_{c}}{y_{t}} \phi_{0}^{2}, \quad 
g \equiv \frac{y_{u}}{y_{t}} \phi_{0}^{\prime 3} \; .
\end{equation} 
With the numerical choice of 
\begin{equation}
b = 0.00304, \quad 
c = -0.0172,  \quad 
k = -0.0266,  \quad 
h = 0.00426, \quad 
g = 1.45\times 10^{-5} \; , 
\end{equation}
the following mass ratios are obtained, 
\begin{equation}
m_{d}: m_{s} : m_{b} \simeq \theta_{c}^{\scriptscriptstyle 4.6} : \theta_{c}^{\scriptscriptstyle 2.7} : 1, \qquad   
m_{u} : m_{c} : m_{t} \simeq  \theta_{c}^{\scriptscriptstyle 7.5} : \theta_{c}^{\scriptscriptstyle 3.7} : 1 \; ,
\end{equation} 
where $\theta_{c} \simeq \sqrt{m_{d}/m_{s}} \simeq 0.225$ is the Cabbibo angle. These ratios in terms of powers of $\theta_{c}$ agree with those given in ~\cite{McKeen:2007ry}. We have also taken 
\begin{equation}
y_{t}/\sin\beta = 1.25, \quad 
y_{b}\phi_{0} \zeta_{0}/\cos\beta \simeq m_{b}/m_{t} \simeq 0.011, \quad  
\tan\beta = 10 \; . 
\end{equation}
The effects of the renormalization group corrections have been taken into account in our numerical predictions, which are quoted at the electroweak scale.  As a result of the GJ relations, realistic charged lepton masses are obtained. 
Making use of these input parameters, the complex CKM matrix is,
\begin{eqnarray}
\left( \begin{array}{ccc}
0.974e^{-i 25.4^{\circ}} & 0.227 e^{i23.1^{\circ}} & 0.00412e^{i166^{\circ}} \\
0.227 e^{i123^{\circ}} & 0.973 e^{-i8.24^{\circ}} & 0.0412 e^{i180^{\circ}} \\
0.00718 e^{i99.7^{\circ}} & 0.0408 e^{-i7.28^{\circ}} & 0.999
\end{array}\right). 
\end{eqnarray}
For the three angles in the unitarity triangle, our model predictions are, 
\begin{equation}
\beta = 23.6^{\circ} \quad (\sin2\beta  =  0.734) \; , \quad
\alpha = 110^{\circ} \; , \quad 
\gamma = \delta_{q} = 45.6^{\circ} \; ,
\end{equation}
(where $\delta_{q}$ is the CP phase in the standard parametrization), and they agree with the direct measurements within $1\sigma$ of BaBar and $2\sigma$ of Belle (M. Antonelli et al in Ref.~\cite{Amsler:2008zzb}). 
Our predictions for the Wolfenstein parameters are, 
\begin{equation}
\lambda=0.227 \; , \quad 
A=0.798 \; , \quad 
\overline{\rho} = 0.299 \; , \quad 
\overline{\eta}=0.306 \; , 
\end{equation}
which are very close to the global fit values except for $\overline{\rho}$. The Jarlskog invariant is predicted to be, 
\begin{equation}
\text{J}  \equiv  \mbox{Im} (V_{ud} V_{cb} V_{ub}^{\ast} V_{cd}^{\ast}) = 2.69 \times 10^{-5} \; ,
\end{equation} 
in the quark sector and also agrees with the current global fit value. Potential direct measurements at the LHCb for these parameters can test our predictions.

\section{Neutrino Masses, Lepton Mixings and CP Phases}
\label{sec:neutrino}
\subsection{Neutrino Masses and Lepton Mixings}

The small neutrino masses are generated through the Type-I see-saw mechanism. The right-handed Majorana mass matrix is given by,
\begin{displaymath}
M_{RR} = \left( \begin{array}{ccc}
1 & 0 & 0 \\
0 & 0 & 1 \\
0 & 1 & 0 
\end{array}\right) s_{0} \Lambda
\end{displaymath}
with the Dirac neutrino mass matrix being,
\begin{displaymath}
M_{D} = \left( \begin{array}{ccc}
2\xi_{0} + \eta_{0} & -\xi_{0} & -\xi_{0} + \eta_{0}^{\prime\prime} \\
-\xi_{0} & 2\xi_{0} + \eta_{0}^{\prime\prime} & -\xi_{0} + \eta_{0} \\
-\xi_{0} + \eta_{0}^{\prime\prime} & -\xi_{0}+\eta_{0} & 2\xi_{0} 
\end{array}\right) \zeta_{0} \zeta^{\prime}_{0} v_{u} \;.
\end{displaymath}
Without loss of generality, we have implicitly set $\lambda_{1, \; 2, \; 3, \; 4} = 1$. 
In the type-I seesaw framework, after integrating out the right-handed neutrinos, the effective neutrino mass matrix is given by,
\begin{equation}
M_{\nu}^{\mbox{\tiny{eff}}} \simeq -M_D
\cdot 
M_{RR}^{-1}
\cdot
M_D^{T} \;  .
\end{equation}
Specifically for our model, the effective neutrino mass matrix is given in terms of the flavon VEVs as,
\begin{eqnarray}
M_{\nu}^{\mbox{\tiny{eff}}} & \simeq & -\frac{(\zeta_0\zeta_0^{\prime}v_u)^2}{s_0\Lambda} \\ \nonumber
& \times & \left(\begin{array}{ccc} 6\xi_0^2 + 4\xi_0\eta_0 + \eta_0^2 -2\xi_0\eta_0^{\prime \prime} & \eta_0^{\prime \prime 2}-3\xi_0^2+\xi_0(\eta_0^{\prime \prime}-2\eta_0) & 2\eta_0\eta_0^{\prime \prime}-3\xi_0^2+\xi_0(\eta_0^{\prime \prime}-2\eta_0) \\ \eta_0^{\prime \prime 2}-3\xi_0^2+\xi_0(\eta_0^{\prime \prime}-2\eta_0) & \xi_0^2+2(\eta_0-\xi_0)(2\xi_0+\eta_0^{\prime \prime}) & 6\xi_0^2+\eta_0^2+\xi_0(\eta_0^{\prime \prime}-2\eta_0) \\ 2\eta_0\eta_0^{\prime \prime}-3\xi_0^2+\xi_0(\eta_0^{\prime \prime}-2\eta_0) & 6\xi_0^2+\eta_0^2+\xi_0(\eta_0^{\prime\prime}-2\eta_0) & 4\xi_0(\eta_0-\xi_0)+(\xi_0-\eta_0^{\prime\prime})^2  
\end{array}\right) 
 \; ,
\end{eqnarray}
where $-\frac{(\zeta_0\zeta_0^{\prime}v_u)^2}{s_0\Lambda}$ is the overall neutrino mass scale. 
In the absence of the contribution from the singlet scalar, $\eta_0^{\prime\prime} = 0$, which corresponds to the exact model in~\cite{Chen:2007afa}, the effective neutrino mass matrix $M^{\nu}_{\mbox{\tiny{eff}}}$ is diagonalized by the TBM mixing matrix. 

With $\eta_0^{\prime\prime} \neq 0$, the neutrino diagonalization matrix deviates from the TBM matrix. Therefore, in order to understand the structure of the diagonalization matrix, in particular the analytic form of the deviations from TBM mixing pattern, we first multiply the effective neutrino mass matrix by $U_{\mbox{\tiny TBM}}$ on both sides. The resulting mass matrix 
$M^{\mbox{\tiny{eff}}}_{\mbox{\tiny{TBM}}} =  
U_{\mbox{\tiny{TBM}}}^T
\cdot
M^{\mbox{\tiny{eff}}}_{\nu}
\cdot
U_{\mbox{\tiny{TBM}}}$ 
is simplified as,
\begin{eqnarray}
\label{MeffTBM}
M^{\mbox{\tiny{eff}}}_{\mbox{\tiny{TBM}}} \simeq  -\frac{(\zeta_0\zeta_0^{\prime}v_u)^2}{s_0\Lambda}\left(\begin{array}{ccc}(3\xi_0+\eta_0-\frac{\eta_0^{\prime \prime}}{2})^2-\frac{3}{4}\eta_0^{\prime \prime 2} & 0 & \frac{\sqrt{3}}{2}(2\eta_0-\eta_0^{\prime \prime})\eta_0^{\prime \prime} \\ 0 & (\eta_0 + \eta_0^{\prime \prime})^2 & 0 \\ \frac{\sqrt{3}}{2}(2\eta_0 - \eta_0^{\prime \prime})\eta_0^{\prime\prime} & 0 & -(3\xi_0-\eta_0+\frac{\eta_0^{\prime\prime}}{2})^2+\frac{3}{4}\eta_0^{\prime\prime 2}\end{array} \right)\;.
\end{eqnarray} 
We immediately find that the resulting mass matrix $M_{\mbox{\tiny{TBM}}}^{\mbox{\tiny{eff}}}$ is diagonalizable by a further rotation in the $(1,3)$-plane 
\begin{equation}
\label{eqn:diagNeuMass}
 Q^{T} \cdot U_{\varphi}^T \cdot M^{\mbox{\tiny{eff}}}_{\mbox{\tiny{TBM}}} \cdot U_{\varphi} \cdot Q 
 = Q^{T} \cdot \text{diag}(m_1, m_2, m_3) \cdot Q 
= \text{diag}(|m_1|, |m_2|, |m_3|) \; ,
\end{equation}
with the rotation matrix $U_{\varphi}$ defined as,
\begin{eqnarray}
\label{mtx:ualpha}
U_{\varphi} = \left(\begin{array}{ccc}\cos\varphi & 0 & -\sin\varphi \\ 0 & 1 & 0 \\ \sin\varphi & 0 & \cos\varphi \end{array}\right) \;,
\end{eqnarray} 
$Q$ being a diagonal phase matrix, and $|m_1|$, $|m_2|$, and $|m_3|$ being the absolute effective neutrinos masses.  
Defining the following variables,
\begin{eqnarray}
-\frac{(\zeta_0\zeta_0^{\prime}v_u)^2}{s_0\Lambda}\left((3\xi_0+\eta_0-\frac{\eta_0^{\prime \prime}}{2})^2-\frac{3\eta_0^{\prime \prime 2}}{4}\right) & \equiv & {\bf a} \;;\\
-\frac{(\zeta_0\zeta_0^{\prime}v_u)^2}{s_0\Lambda}\left(\sqrt{3}(2\eta_0-\eta_0^{\prime \prime})\frac{\eta_0^{\prime \prime}}{2}\right) & \equiv & {\bf b} \;; \\
-\frac{(\zeta_0\zeta_0^{\prime}v_u)^2}{s_0\Lambda}\left(-(3\xi_0-\eta_0+\frac{\eta_0^{\prime\prime}}{2})^2+\frac{3\eta_0^{\prime\prime 2}}{4}\right) & \equiv & {\bf c} \;,
\end{eqnarray}
the neutrinos masses $m_1,\; m_3$ and the additional rotational angle $\varphi$ can then be written as functions of {\bf a}, {\bf b} and {\bf c}. We find
\begin{eqnarray}
m_1 & = & \left(\frac{{\bf a}+{\bf c}}{2} + \frac{\sqrt{({\bf a}-{\bf c})^2+4{\bf b}^2}}{2}\right) \;,\\
m_2 & = & (\eta_0 + \eta_0^{\prime \prime})^2\;,\\
m_3 & = & \left(\frac{{\bf a}+{\bf c}}{2} - \frac{\sqrt{({\bf a}-{\bf c})^2+4{\bf b}^2}}{2}\right)  \;,\\
\sin2\varphi & = & \frac{2{\bf b}}{\sqrt{({\bf a}-{\bf c})^2+4{\bf b}^2}} \;, \\
\cos2\varphi & = & \frac{{\bf a}-{\bf c}}{\sqrt{({\bf a}-{\bf c})^2+4{\bf b}^2}} \;.
\end{eqnarray} 

As the overall phase for all masses is not physical, without losing any generality, we choose the overall mass scale $\frac{(\zeta_0\zeta_0^{\prime}v_u)^2}{s_0\Lambda}$ to be negative so that $m_2$ is positive. For ${\bf a} + {\bf c} < 0$, normal mass hierarchy is predicted;  otherwise, it is inverted mass hierarchy. Furthermore, $m_1$ is always positive unless both ${\bf a} + {\bf c} < 0$ and ${\bf a}{\bf c} > {\bf b}^2$ are satisfied. Similarly, $m_3$ is always negative unless ${\bf a} + {\bf c} > 0$ and ${\bf a}{\bf c} > {\bf b}^2$ are satisfied simultaneously.
The phase matrix, $Q$ is given by $Q=\text{diag}(1, 1, e^{i \pi})$ for inverted hierarchy, and $Q=\text{diag}(e^{i \pi}, 1, e^{i \pi})=(e^{i \pi})\text{diag}(1, e^{- i \pi}, 1)$ for the normal hierarchy. The overall coefficient $(e^{i \pi})$ for the normal hierarchy case is an overall phase and has no physical meaning. 

There are additional mass sum rules such as,
\begin{equation}
m_1 + m_2 + m_3 
= - \biggl[ 
12\xi_0 
 \biggl(
\eta_0-\frac{\eta_0^{\prime\prime}}{2}
 \biggl)
+(\eta_0+\eta_0^{\prime\prime})^2
 \biggr]
\frac{(\zeta_0\zeta_0^{\prime}v_u)^2}{s_0\Lambda} \;,
\end{equation}
and in terms of the absolute neutrino masses,
\begin{eqnarray}
& &-\frac{|m_1|+|m_2|+|m_3|}{\frac{(\zeta_0\zeta_0^{\prime}v_u)^2}{s_0\Lambda} } =\\
& &\left\{\begin{array}{ll} 12\xi_0(\eta_0-\frac{\eta_0^{\prime\prime}}{2})+(\eta_0+\eta_0^{\prime\prime})^2 \; , 
(m_1 > 0, \; m_3 > 0) ; 
\vspace{0.2in} \\ 
2\sqrt{81\xi_0^4+9\xi_0^2(2\eta_0^2-2\eta_0\eta_0^{\prime\prime}-\eta_0^{\prime\prime 2})+(\eta_0^2-\eta_0\eta_0^{\prime \prime} + \eta_0^{\prime\prime   2})^2}+(\eta_0+\eta_0^{\prime\prime})^2 \; ,    (m_1 > 0 , \; m_3 < 0) ;
\vspace{0.2in} \\ 
-12\xi_0(\eta_0-\frac{\eta_0^{\prime\prime}}{2})+(\eta_0+\eta_0^{\prime\prime})^2 \;  , 
 (m_1 < 0, \; m_3 < 0 ). \end{array} \right. \
\nonumber
\end{eqnarray}
which can be used to check against our numerical results. It is worth noting that some previously found sum rules occur in models with two free parameters in the neutrino sector. In this case,  one of the masses is constrained against the other two. Our model features three free parameters, which in turn constrains the sum of the masses to be some combination in terms of the model parameters as shown in eq. 36. For examples of sum rules in models with only two parameters in the neutrino sector, see~\cite{Barry:2010yk,Dorame:2011eb}.

The full diagonalization matrix in the neutrino sector is given by $U_{\nu}^D = U_{\mbox{\tiny{TBM}}} \cdot U_{\varphi} \cdot Q$, 
\begin{eqnarray}
U_{\nu}^D = \left(\begin{array}{ccc}\sqrt{\frac{2}{3}}\cos\varphi & \sqrt{\frac{1}{3}} & -\sqrt{\frac{2}{3}}\sin\varphi \\ -\sqrt{\frac{1}{6}}\cos\varphi -\sqrt{\frac{1}{2}}\sin\varphi & \sqrt{\frac{1}{3}} & \sqrt{\frac{1}{6}}\sin\varphi-\sqrt{\frac{1}{2}}\cos\varphi \\ -\sqrt{\frac{1}{6}}\cos\varphi + \sqrt{\frac{1}{2}}\sin\varphi & \sqrt{\frac{1}{3}} & \sqrt{\frac{1}{6}}\sin\varphi + \sqrt{\frac{1}{2}}\cos\varphi \end{array}\right) \cdot Q \;\;\;  .
\end{eqnarray}
The corresponding mixing angles are then given by,
\begin{eqnarray}
\tan^2\theta_{13}^{\nu} & = & \frac{1-\cos 2\varphi}{2+\cos 2\varphi}\;; \\
\tan^2\theta_{12}^{\nu} & = &\frac{1}{1+\cos 2\varphi}\;;\\
\tan^2\theta_{23}^{\nu} & = &  \frac{2+\cos 2\varphi-\sqrt{3}\sin 2\varphi}{2+\cos 2\varphi+\sqrt{3}\sin 2\varphi} \; .
\end{eqnarray}
In the limit of $\varphi = 0$, these expressions coincide with the TBM predictions, 
\begin{equation}
\sin^{2} \theta_{23} = 1/2, \quad \sin^{2} \theta_{12} = 1/3, \quad \sin\theta_{13} = 0 \; ,
\end{equation}
as expected.

To obtain the full leptonic mixing matrix, the diagonalization matrix in the charged lepton sector must be taken into account. Given the $SU(5)$ relation, $M_{e} = M_{d}^{\mbox{\tiny T}}$ (up to the $SU(5)$ CG's) in our model, the charged lepton mixing matrix is strongly constrained by the quark sector.  
The charged lepton mixing matrix $U_{e,L}$ is fixed by the $SU(5)$ symmetry to be, 
\begin{equation}
\label{eqn:uelnum}
U_{e,L} = \left( \begin{array}{ccc}
0.997 e^{i177^{\circ}} & 0.0823 e^{i131^{\circ}} & 1.31\times10^{-5} e^{-i45^{\circ}}  \\
0.0823 e^{i41.8^{\circ}} & 0.997 e^{i176^{circ}} & 0.000149 e^{-i3.58^{\circ}}  \\
1.14\times10^{-6} & 0.000149 & 1
\end{array}\right) \;,
\end{equation}
Up to field rephasing, the diagonalization matrix $U_{e,L}$ can be approximated, at the leading order, as
\begin{equation}
U_{e,L} \simeq 
\left( \begin{array}{ccc}
1 & \theta_{c} e^{-i\delta_{e}} /3 & 0 \\
\theta_{c} e^{i\delta_{e}}/3 & 1 & 0 \\
0 & 0 & 1
\end{array}
\right) \; .
\end{equation}

By multiplying the neutrino mixing matrix $U_{\nu}^D$, we obtain the full lepton mixing matrix $U^{\ell} = U_{e,L}^{\dagger} \cdot U_{\nu}^D$, which is given in the standard parametrization as follows,
\begin{equation}
U_{\mbox{\tiny{PMNS}}} = \left(\begin{array}{ccc}
c_{12}c_{13} & s_{12}c_{13} & s_{13}e^{-i\delta} \\
-s_{12}c_{23}-c_{12}s_{23}s_{13}e^{i\delta} & c_{12}c_{23}-s_{12}s_{23}s_{13}e^{i\delta} & s_{23}c_{13} \\
s_{12}s_{23}-c_{12}c_{23}s_{13}e^{i\delta} & -c_{12}s_{23}-s_{12}c_{23}s_{13}e^{i\delta} & c_{23}c_{13}
\end{array}\right) \left(\begin{array}{ccc}1 & 0 & 0 \\ 0 & e^{i\alpha_1/2} & 0 \\ 0 & 0 & e^{i\alpha_2/2} \end{array}\right) \; , 
\end{equation}
where $c_{ij} \equiv \cos\theta_{ij},\; s_{ij} \equiv \sin\theta_{ij}$ are the lepton mixing angles, $\delta$ is the Dirac CP phase, and $\alpha_1,\; \alpha_2$ are the Majorana CP phases. 

To the first order, the lepton mixing angle $\theta_{13}$ can be approximated by, 
\begin{eqnarray}
U_{e3} = \sin\theta_{13} e^{-i\delta}
& \simeq & 
\frac{\theta_{c}e^{-i\delta_{e}}}{3\sqrt{2}} \cos \varphi  
+ \biggl[ \theta_{13}^{\nu} + \kappa \frac{\theta_{c}e^{-i\delta_{e}}}{3} \biggr] 
\\
& = & 
\frac{\theta_{c}e^{-i\delta_{e}}}{3\sqrt{2}} \cos \varphi  
+ \biggl[ -\sqrt{\frac{2}{3}} + \frac{\theta_{c}e^{-i\delta_{e}}}{3\sqrt{6}} \biggr] \sin\varphi
\nonumber
\; ,
\end{eqnarray}
where $\theta_{c}/3\sqrt{2}$ arises from the charged lepton sector while $\theta_{13}^{\nu} = -\sqrt{2/3} \sin\varphi$ and $\kappa = \sqrt{1/6} \sin\varphi$ are contributions from the neutrino sector. The later two terms are the additional contributions due to the additional singlet $\eta^{\prime\prime}$ in our present modified model. For small $\varphi$, an expansion around $\varphi = 0$ leads to the following analytic expressions for $\theta_{13}$ and $\delta$:
\begin{eqnarray}
\theta_{13} & \simeq & |U_{e3}| \simeq \frac{\theta_{c}}{3\sqrt{2}} + \varphi \biggl[ \frac{\theta_{c}-6\cos(\delta_e)}{3\sqrt{6}} \biggr] + \mathcal{O}(\varphi^{2},\theta_c^2)  \; ,
\\
\delta & = & - \text{Arg} \biggl[ U_{e3} \biggr] +\pi \simeq \delta_{e} + \varphi \biggl[ \frac{2\sqrt{3}\sin(\delta_{e})}{\theta_{c}}  \biggr] + \pi + \mathcal{O}(\varphi^{2},\theta_c^2)   \; .
\end{eqnarray}
Similarly, the sum rule for the solar mixing angle $\theta_{12}$ and $\theta_{c}$ is found to be, 
\begin{equation}
\tan\theta^{2}_{12} \simeq \frac{1}{2} + \frac{\theta_{c}}{3} \cos(\delta_{e}) +  \varphi\biggl[  \frac{\theta_{c}\cos(\delta_{e})}{2\sqrt{3}} \biggr] + \mathcal{O} (\varphi^{2},\theta_c^2) \; .
\end{equation}
With the full charged lepton diagonalization matrix $U_{e,L}$ given in Eq.~(\ref{eqn:uelnum}), the full lepton mixing matrix $U^{\ell} = U_{e,L}^{\dagger} \cdot U_{\nu}^D$ 
can be expanded as a function of $c_{\varphi} \equiv \cos\varphi$ and $s_{\varphi} \equiv \sin\varphi$,
\begin{eqnarray}
U^{\ell} \simeq \left(\begin{array}{lcl}
-(0.838+0.0202i) c_\varphi & (-0.539-0.0618i) & -(0.0434-0.0388i) c_\varphi \\
\quad\quad-(0.0434-0.0388i) s_\varphi & & \quad\quad+(0.838+0.0202i) s_\varphi\\
 (0.362-0.0223i) c_\varphi & (-0.605-0.0760i) & (0.703+0.0492i) c_\varphi \\
\quad\quad+ (0.703+0.0492i) s_\varphi & & \quad\quad-(0.362-0.0223i) s_\varphi\\
 -0.408c_\varphi + 0.707s_\varphi & 0.577 & 0.707c_\varphi + 0.408 s_\varphi\end{array}\right)\cdot Q\;.
\end{eqnarray}
The lepton mixing angles can then be expressed as the following,
\begin{eqnarray}
\tan^2\theta_{13}^{\ell} & \simeq & \frac{1.010-1.001c_{2\varphi} - 0.102 s_{2\varphi}}{1.853+c_{2\varphi}+0.101s_{2\varphi}}\;; \\
\tan^2\theta_{12}^{\ell} & \simeq &\frac{0.843}{1.010+c_{2\varphi} + 0.102s_{2\varphi}}\;;\\
\tan^2\theta_{23}^{\ell} & \simeq &\frac{1.887+1.098c_{2\varphi} - 1.522s_{2\varphi}}{2.001+c_{2\varphi}+1.733s_{2\varphi}}  \;.
\end{eqnarray}
where 
$c_{2\varphi} \equiv \cos2\varphi$ and $s_{2\varphi} \equiv \sin2\varphi$. 
By applying the various constraints from the mixing angles summarized in Table.~(\ref{tbl:exp_results}) up to $2\sigma$ confidence level,
 we obtain the following range of allowed values for the the angle $\varphi$. For the normal mass hierarchy (NH) and inverted mass hierarchy (IH), the range of allowed values for $\tan 2\varphi$ are, respectively,
\begin{equation}
\label{eqn:range}
-0.337 \lesssim \tan2\varphi \lesssim -0.218 (\text{NH});\quad -0.344 \lesssim \tan2\varphi \lesssim  -0.228 (\text{IH}) \; .
\end{equation}
These correspond to the following values for the angle $\varphi$,
\begin{equation}
 -9.305^{\circ}  \lesssim \varphi \lesssim-6.155^{\circ}  (\text{NH});\quad -9.501^{\circ} \lesssim \varphi \lesssim -6.414^{\circ} (\text{IH})\;.
\end{equation}

\begin{figure}[b]
\centering
\subfloat{\includegraphics[scale=0.25]{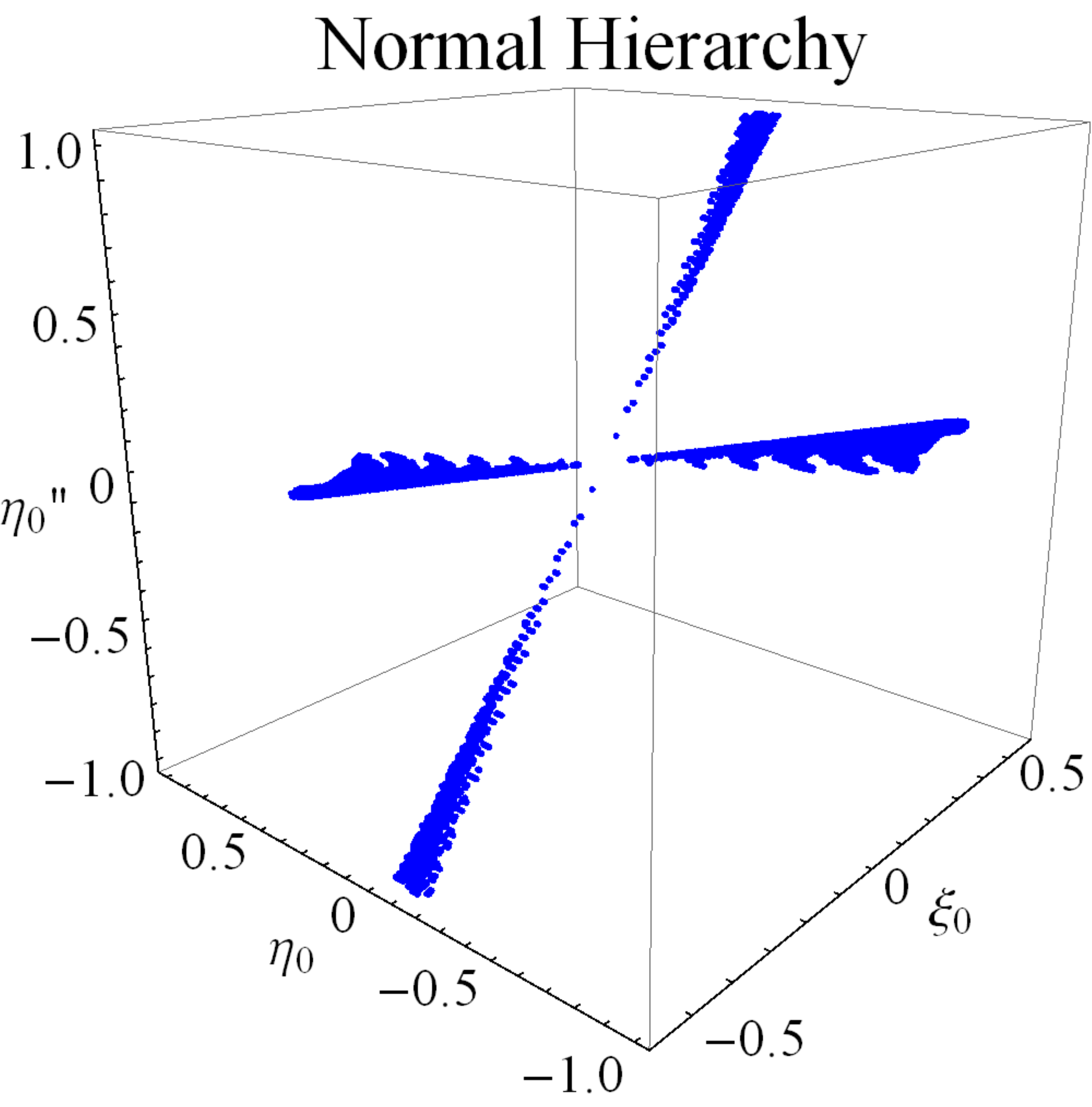}\label{fig:normHregion}} $\quad$
\subfloat{\includegraphics[scale=0.25]{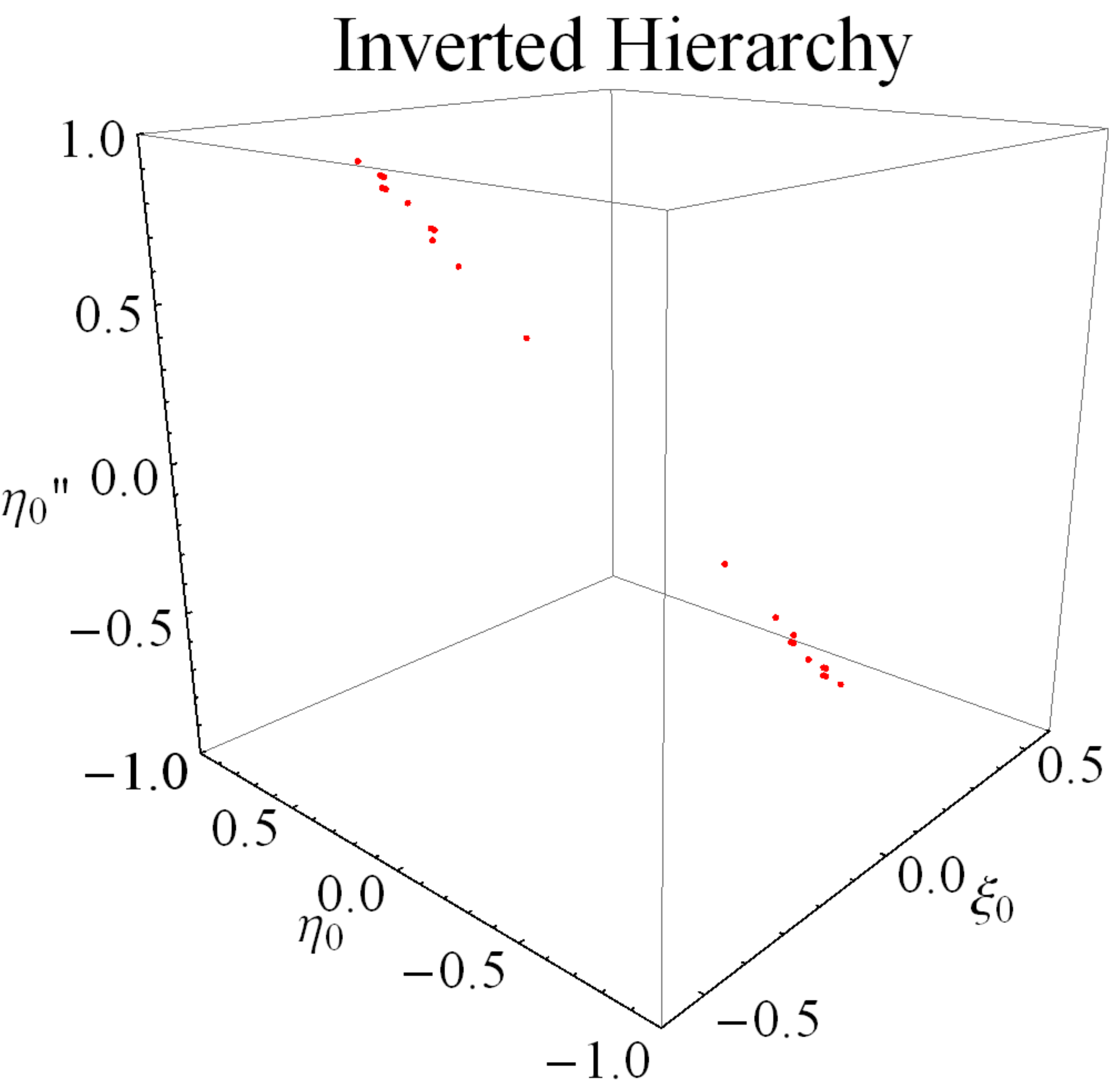}\label{fig:invtHregion}}
	\caption{The allowed parameter spaces of the neutrino flavon VEVs ($\xi_0,\; \eta_0,\; \eta_0^{\prime \prime}$) consistent with the global fits of the lepton mixing angles and the mass squared splittings within the $2\sigma$ range. The left and right panel correspond to the normal neutrino mass hierarchy and inverted neutrino mass hierarchy respectively.}
	\label{fig:region}
\end{figure}

We scan the parameter space of the neutrino flavon VEVs ($\xi_0,\; \eta_0,\; \eta_0^{\prime \prime}$), and find a volume satisfying the lepton mixing angles and the mass squared splittings up to the $2\sigma$ accuracy. These allowed regions are shown in the following two figures (Fig.~\ref{fig:region}). 
The left panel corresponds to the normal mass hierarchy while the right one illustrates the allowed range for the inverted mass hierarchy. Therefore, both NH and IH are allowed in this model, which is different from the prediction in the model~\cite{Chen:2007afa} where only NH exists. However, from the Fig.~\ref{fig:region}, we notice that it is much easier to satisfy all of the experimental observations in the NH than the IH in this model. More interestingly, we notice that roughly the $\eta_0^{\prime\prime}$ parameter depends on the parameter $\xi_0$ linearly when $\eta_0 \ll 1$ and it is due to the constraints on the $\varphi$ rotational angle. It can be understood from the relation $\tan 2\varphi = (\sqrt{3}(2\eta_0-\eta_0^{\prime\prime})\eta_0^{\prime\prime})/(18\xi_0^2+2\eta_0^2-2\eta_0\eta_0^{\prime\prime}-\eta_0^{\prime\prime 2})$ while $\eta_0 \rightarrow 0$, which means,
\begin{equation}
(18\tan2\varphi) \xi_0^2 = (\tan2\varphi-\sqrt{3})\eta_0^{\prime\prime 2} \;.
\end{equation}
We can notice the linear relationship between $\xi_0$ and $\eta_0^{\prime\prime}$ immediately with a fixed $\tan2\varphi$, which is consistent with the Eq.~(\ref{eqn:range}) where $\tan2\varphi$ is close to a constant. The other section shown in FIG. \ref{fig:region} can be understood as a conical section with an elliptical projection. For fixed $\varphi$, one can also find the relation:
\begin{equation}
18\tan(2\varphi) \xi_0^2+\frac{3}{4}\left(\sqrt{3}-\frac{4}{\sqrt{3}-\tan(2\varphi)}+3\tan(2\varphi)\right)\eta_0^2=\left(\tan(2\varphi)-\sqrt{3}\right)\left(\eta_0^{\prime\prime}+\frac{\tan(2\varphi)+\sqrt{3}}{2(\tan(2\varphi)-\sqrt{3})}\eta_0\right)^2 \; .
\end{equation}
Restrictions on these surfaces are determined by mass splitting constraints.

\begin{figure}[t!]
\centering
\subfloat{\includegraphics[scale=0.2]{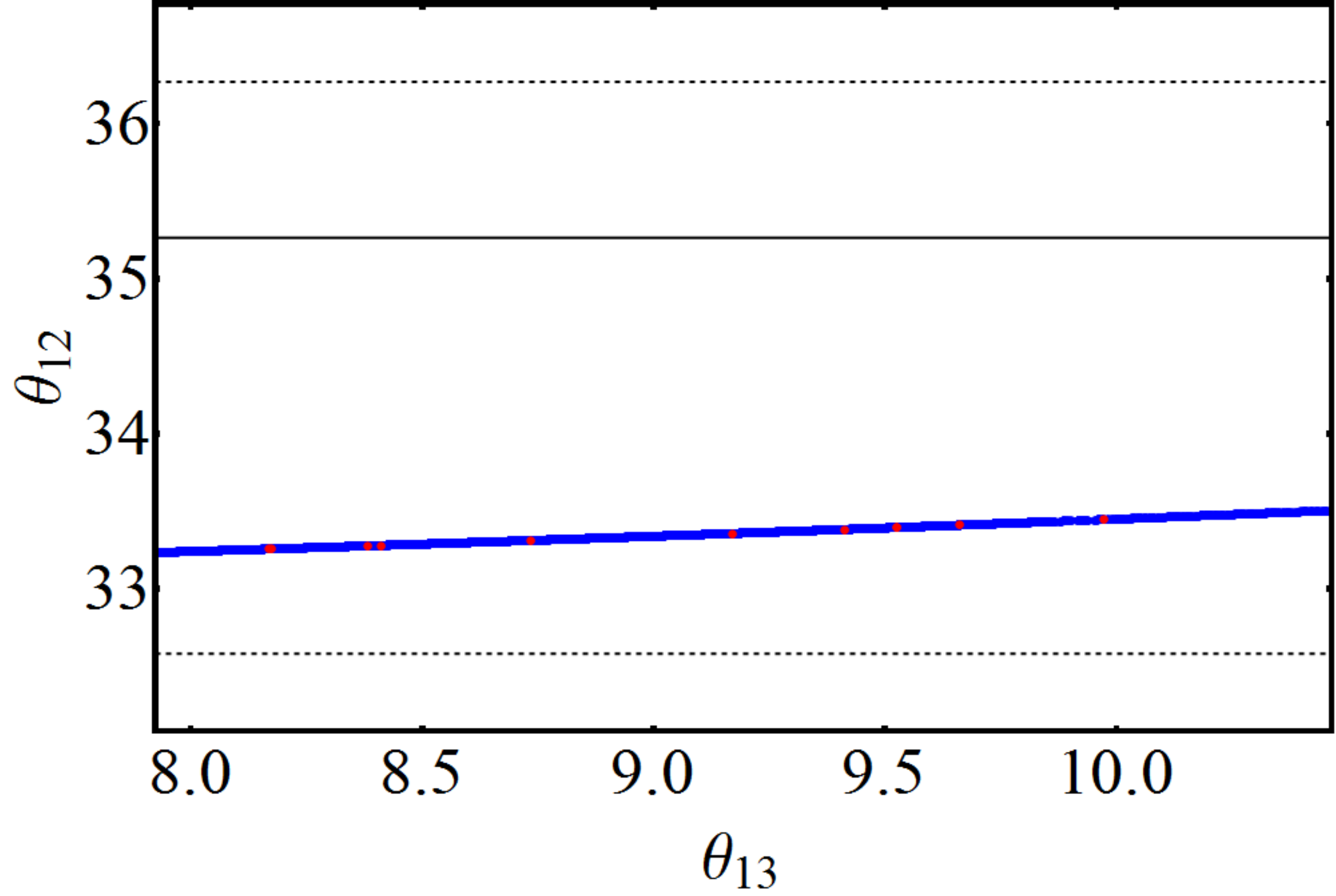}\label{fig:theta12-13}} $\qquad$
\subfloat{\includegraphics[scale=0.2]{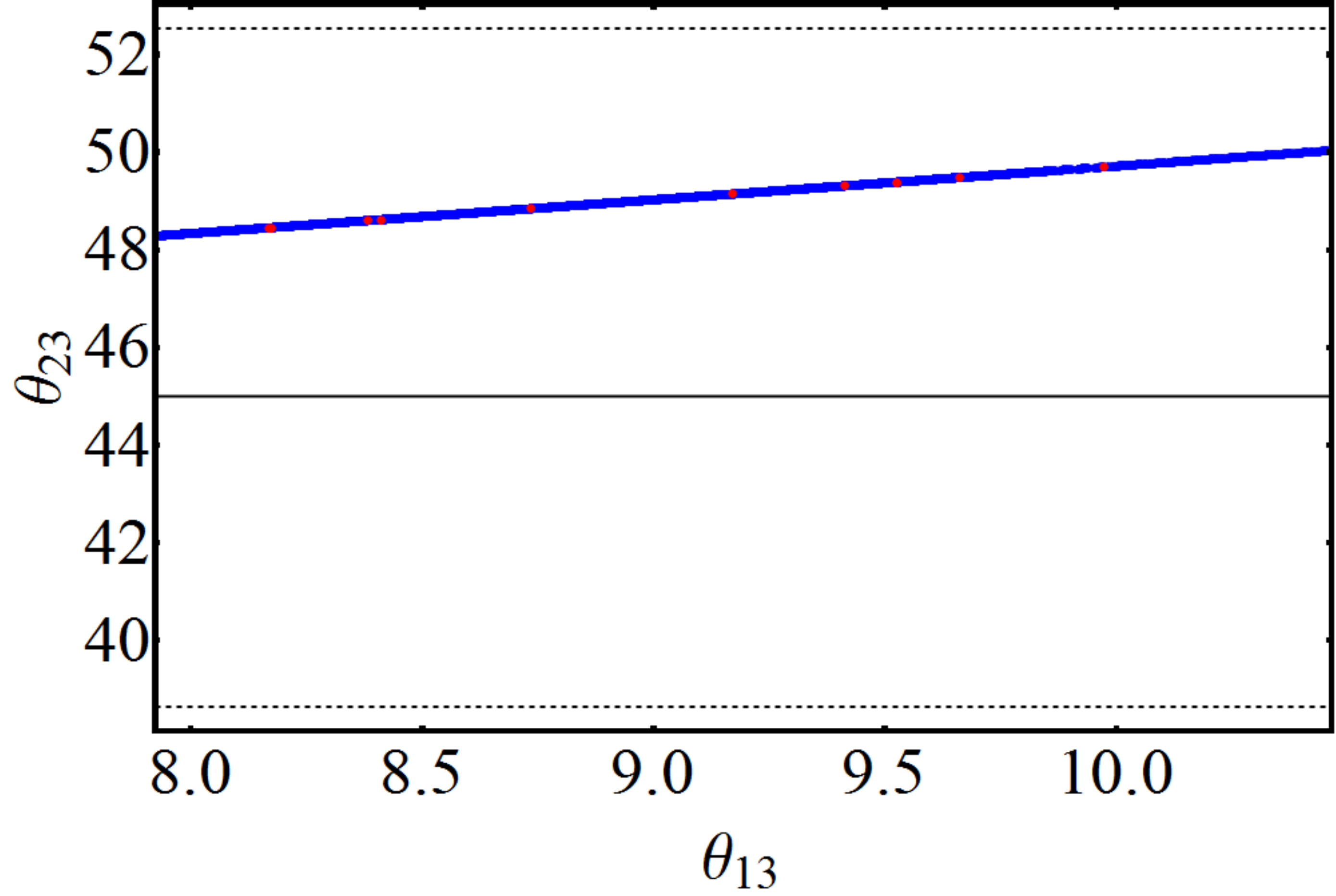}\label{fig:theta23-13}}
	\caption{The distribution of the deviation from TBM mixing vs $\theta_{13}$ this model. Points in the left panel are $\theta_{12}$ vs. $\theta_{13}$ and the points in the right panel are $\theta_{23}$ vs $\theta_{13}$, where blue a blue point indicates a NH solution and a red point indicates an IH solution. The solid black lines correspond to $35.26^{\circ}$ and $45^{\circ}$ and the dotted lines are the $2\sigma$ limits on $\theta_{12}$ and $\theta_{23}$ from the global fit in the left and right panel respectively.}
	\label{fig:diffNH}
\end{figure}

Given that, with a fixed $U_{e,L}$, all three leptonic mixing angles and the Dirac phase are determined by one variable, $\varphi$, there exist correlations among the three mixing angles and phase. The correlations are almost identical for NH and IH, and we show the results found in the parameter scan in Fig.~\ref{fig:diffNH}. Specifically, Fig.~\ref{fig:diffNH} shows the correlations between $\theta_{12}$ and $\theta_{23}$ versus $\theta_{13}$.

In Fig~\ref{fig:diffNH}, the blue and red point in the left and right panels demonstrate the relations between the mixing angle $\theta_{12}$, $\theta_{23}$ and $\theta_{13}$ respectively. Blue (red) points represent the a normal (inverted) hierarchy solution. The solid black lines in the left and right panels correspond to $35.26^{\circ}$ and $45^{\circ}$ respectively while the dotted line are the 2$\sigma$ limits on the mixing angles $\theta_{12}$ and $\theta_{23}$ from the global fit. From Fig.~\ref{fig:diffNH}, we find that for both NH and IH, the mixing angle $\theta_{12}$ tends to be smaller than $35.26^{\circ}$ and as the mixing angle $\theta_{13}$ increases, the deviation decreases, even though the variation in $\theta_{12}$ is very suppressed. However, the trend of the mixing angle $\theta_{23}$ behaves in an opposite way, and it is higher than $45^{\circ}$ and the deviation increases as $\theta_{13}$ is bigger.

Among the large parameter space, we illustrate one numerical example with minimal $\chi^2 = 2.643$. The VEVs in the neutrino sector are fixed to be,
\begin{equation}
\xi_0 = 0.185;\; \quad \eta_0 = -0.865;\; \quad  \eta_0^{\prime \prime} = 0.200\;,
\end{equation}  
leading to the following leptonic mixing angles, 
\begin{equation}
\theta_{12} =  33.351^{\circ}, \;  \quad
\theta_{23} =  49.162^{\circ}, \;  \quad
\theta_{13} =  9.189^{\circ}.
\end{equation}
In addition, we choose the overall mass scale of the neutrinos, $-\frac{(\zeta_0\zeta_0^{\prime}v_u)^2}{s_0\Lambda} = 0.0217\ \text{eV}$, and therefore the neutrino masses are,
\begin{equation}
m_1 = 0.00398\ \text{eV},  \; \quad m_2 = 0.00959\ \text{eV}, \; \quad m_3 = -0.0505\ \text{eV}\;,
\end{equation}
corresponding to normal neutrino mass hierarchy with,
\begin{equation}
\Delta m_{atm}^{2} = 2.53 \times 10^{-3} \; \mbox{eV}^{2}, \quad \quad 
\Delta m_{\odot}^{2} = 7.62 \times 10^{-5} \; \mbox{eV}^{2} \; .
\end{equation}
So all mixing angles and mass squared splittings are consistent with the experimental measurements in the $2\sigma$ region.
Furthermore, the neutrino mass sums are,
\begin{equation}
|m_1| + |m_2| + |m_3| = 0.0640 \; \text{eV}\;, \quad m_1+m_2+m_3 = -0.0369 \; \text{eV}\;.
\end{equation}
We find that our numerical results for the neutrino masses, mixing angles as well as the mass sum rules are consistent with the analytical expressions derived above.

\subsection{Lepton Dirac CP Phase $\delta$, Jarlskog Invariant $J_{\ell}$, Neutrinoless Double Beta Decay Matrix Element $\left<m_{\beta\beta}\right>$}
There are still many fundamental properties of the neutrinos that are unknown at present. These include the Dirac CP phase, the two Majorana phases (if neutrinos are Majorana fermions), the Dirac versus Majorana nature, and the mass hierarchy. With the large value for the $\theta_{13}$ lepton mixing angle, the measurement of the Dirac CP phase is attainable in the future. The parametrization independent CP violation measure, the Jarlskog invariant $J_{\ell}$, in the lepton sector, is related to the Dirac CP phase and is given by,
\begin{equation}
J_{\ell} \equiv \text{Im}[U^{\ell}_{\mu3}U^{\ell*}_{e3}U^{\ell}_{e2}U^{\ell*}_{\mu2}] = 3.338\times10^{-5}-0.00967c_{2\varphi}+0.00554s_{2\varphi}\;.
\end{equation}
$J_{\ell}$ is a function of variable $\varphi$, our additional rotational matrix in the neutrino sector.  Within the range that satisfies the mixing angle measurements at 2$\sigma$ level we derived above, the Jarlskog invariant $\text{J}_{\ell}$ for both NH and IH is restricted to the region,
\begin{equation}
\label{eqn:Jell}
-0.0109 \lesssim J_{\ell} \lesssim -0.0106 \; (\text{NH}/\text{IH}).
\end{equation}
From the Jarlskog invariant $J_{\ell}$, we can further determine the Dirac CP phase. Comparing with the original Dirac CP phase derived in~\cite{Chen:2011vd}, the additional rotational matrix $U_{\varphi}$ does not modify the Dirac CP phase significantly, especially our $\varphi$ rotational angle is small, therefore, the Dirac CP phase is still expected to be close to $5/4\pi\;(225^{\circ})$. Including the effects from the $\varphi$ rotational angle, within the $2\sigma$ allowed range for $\varphi$, the Dirac CP phase can be obtained via $J_{\ell}=(\sin(2\theta_{13})\sin(2\theta_{23})\sin(2\theta_{12})\cos\theta_{13}\sin\delta)/8$ and we find it is restricted to,
\begin{equation}
\label{eqn:dirac}
195.900^{\circ} \lesssim \delta \lesssim 200.139^{\circ} \; (\text{NH}); \quad 195.703^{\circ} \lesssim \delta \lesssim 199.692^{\circ} \;(\text{IH}).
\end{equation} 
As this model assumes neutrinos are Majorana fermions, in addition to the Dirac CP phase there are two additional Majorana CP phases. One way to test it is through the neutrinoless double beta decay ($(\beta\beta)_{0\nu}$-decay). The rate is proportional to the variable, neutrinoless double beta decay matrix element defined as,
\begin{equation}
\left<m_{\beta \beta}\right> = \biggl| 
\sum_{i=1}^3 (U^{\ell}_{ei})^2|m_i|
\biggr|
\;.
\end{equation} 
In this model, it can be written as, 
\begin{eqnarray}
\label{eqn:mbetabeta}
\left<m_{\beta\beta}\right> 
& \simeq & \biggl|(f_1 + f_2c_{2\varphi}+f_3s_{2\varphi})|m_1| + f_4 |m_2| + (f_1-f_2c_{2\varphi}-f_3s_{2\varphi}) |m_3| \biggr|
\end{eqnarray}  
with the following coefficients defined as,  
\begin{eqnarray}
f_1 & = & (0.351+0.0153i);\;\; \quad f_2 = (0.351+0.0186i);\\ \nonumber
f_3 & = & (0.0371-0.0316i);\; \quad f_4 = (0.287-0.0667i).
\end{eqnarray}
From the Eq.~(\ref{eqn:mbetabeta}), the coefficients in front of the masses $m_i\;(i=1,2,3)$ are constants or close to constants due to the narrow allowed region on the rotational angle $\varphi$, therefore, a nearly linear dependence on the neutrinos masses is expected.
\begin{figure}[t!]
\centering
\subfloat{\includegraphics[width=0.5\textwidth]{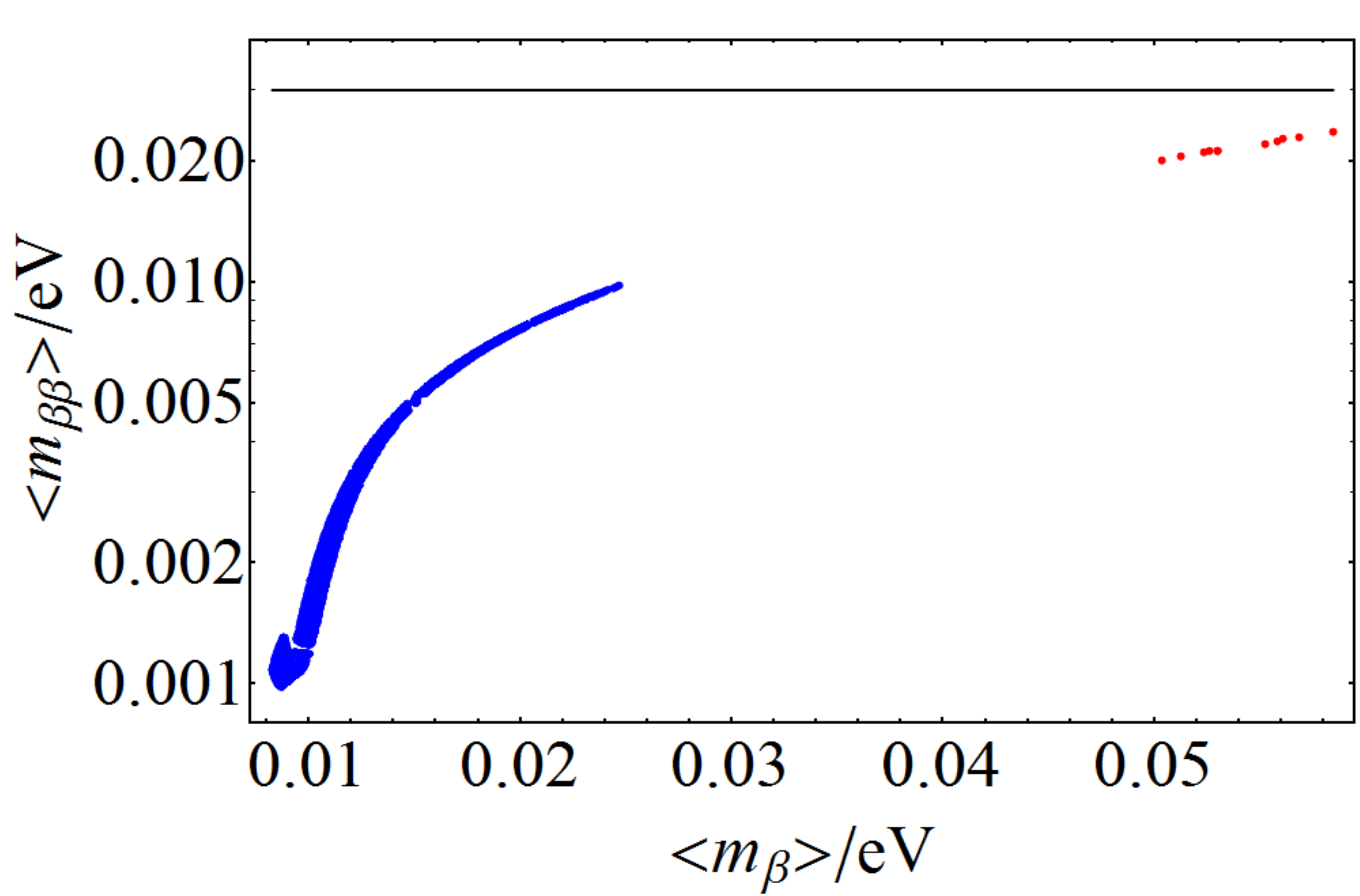}\label{fig:a}}
\subfloat{\includegraphics[width=0.5\textwidth]{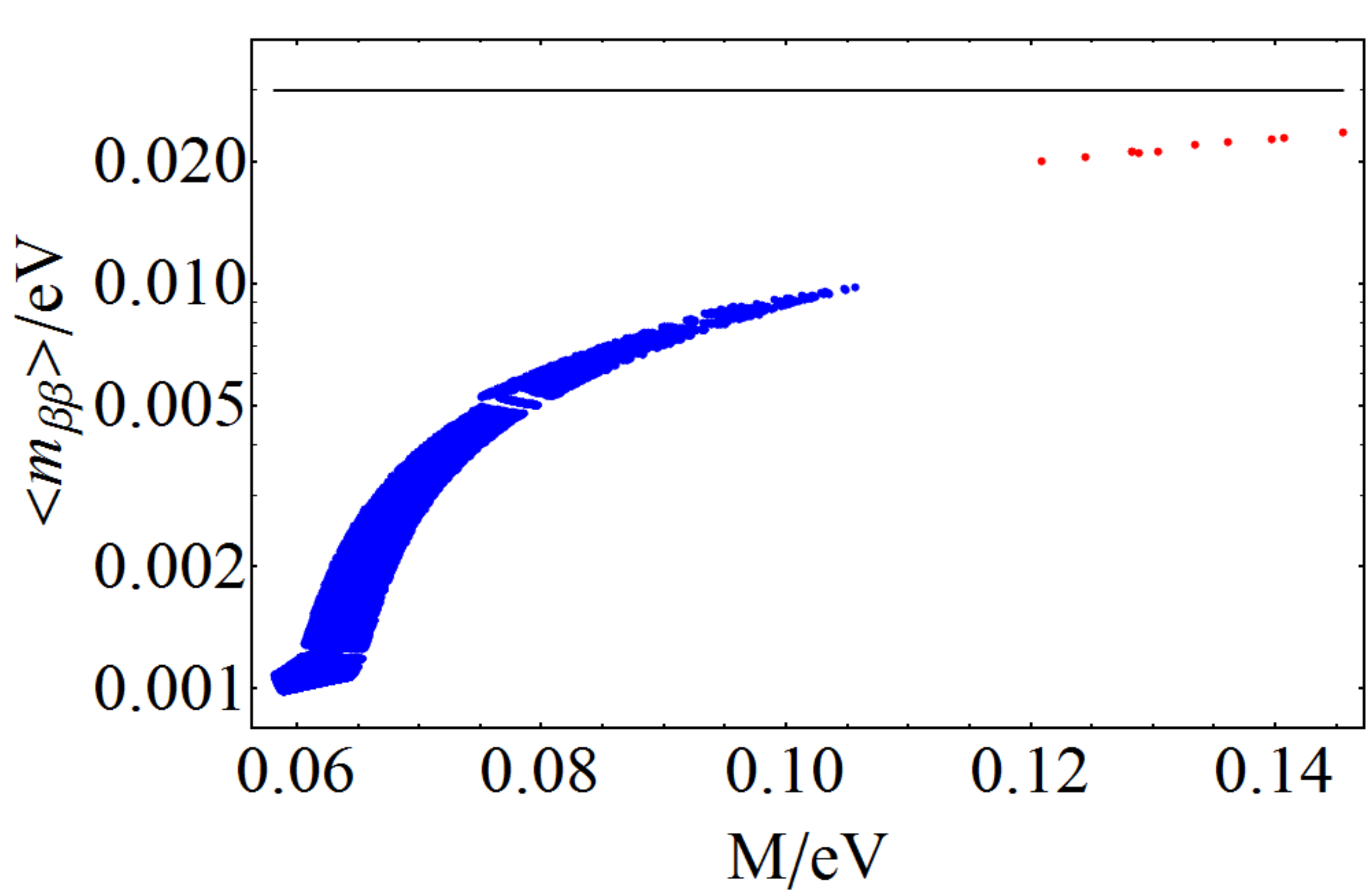}\label{fig:b}} \\
\includegraphics[width=0.5\textwidth]{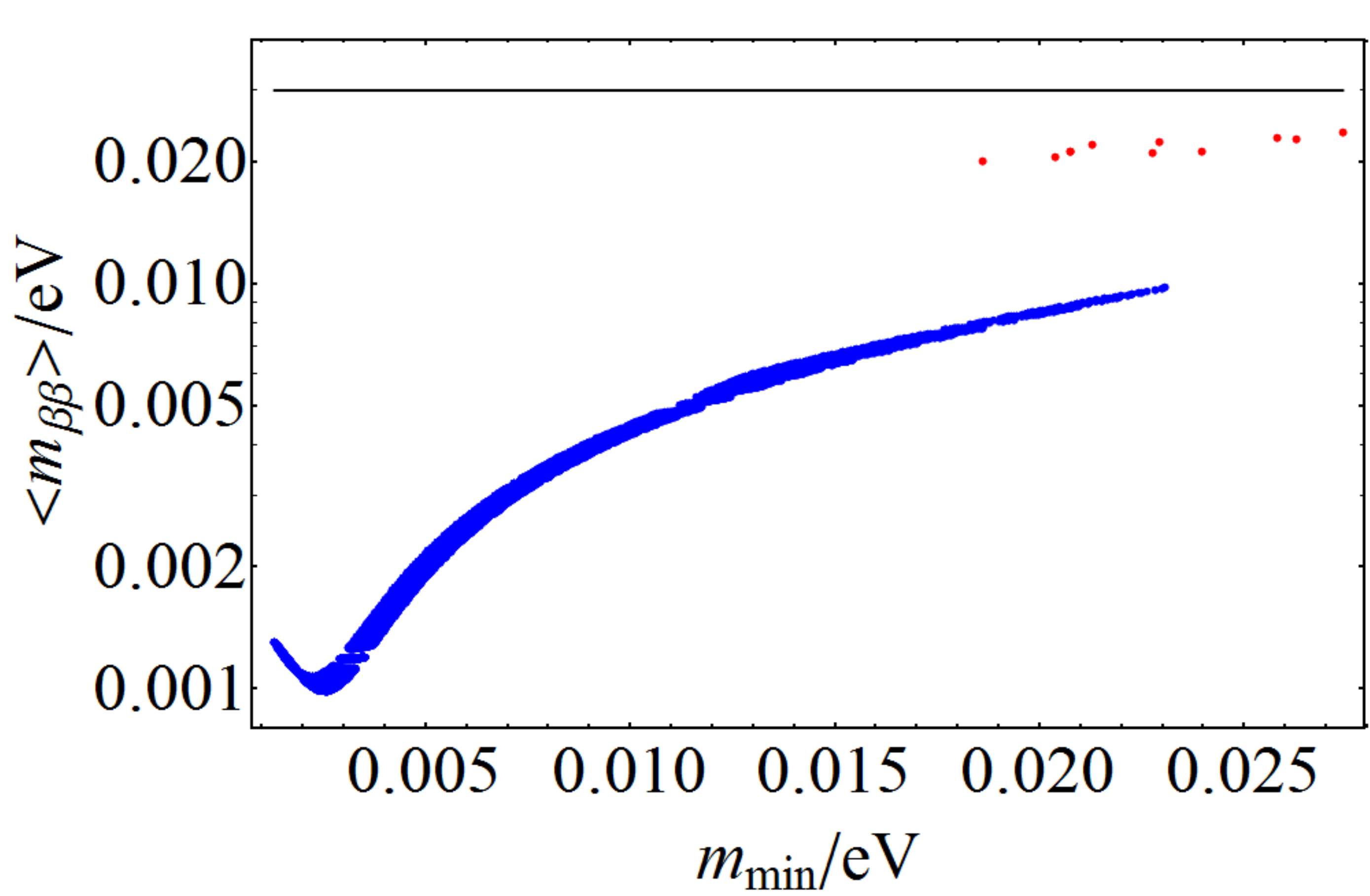}\label{fig:c}
	\caption{
	The neutrinoless double beta decay matrix element verse $\left<m_\beta\right>\equiv\sqrt{\sum_{i=1}^3 |U_{e i}|^2|m_i|^2}$, $\text{M} \equiv |m_1| + |m_2| + |m_3|$ and $m_{\mbox{\tiny{min}}} \equiv \text{min} (|m_{1}|, |m_{2}|, |m_{3}|)$. The blue points are predictions for the normal mass hierarchy, while the red points are predictions for the inverted hierarchy. The black line is the projected reach of the Majorana Demonstrator~\cite{xu:2013}.}
	\label{fig:mbetabeta}
\end{figure}
Shown in Fig.~\ref{fig:mbetabeta} is the neutrinoless double beta decay matrix element $\left< m_{\beta\beta}\right>$ for the normal mass hierarchy, and each point corresponds to the data point in the left panel of Fig.~\ref{fig:region} which satisfies all neutrino mixing angles and mass squared splittings experimentally. Note that the variable $\left<m_{\beta\beta}\right>$ roughly depends on the neutrino masses linearly. In addition, we realize there are many data points with $\left<m_{\beta\beta}\right> \sim 10^{-2}$ eV, which is still below the sensitivity range of various neutrinoless double beta decay experiments~\cite{Branco:2011zb}. The black solid line in the Fig.~\ref{fig:mbetabeta} represents the experimental reach of $\left<m_{\beta\beta}\right> \simeq 3 \times 10^{-2}$ eV with 3 year data at the Majorana Demonstrate~\cite{xu:2013}.

One specific example with minimal $\chi^2$ predicts the Dirac CP phase to be,
\begin{equation}
\delta = 197.734^{\circ} \;.
\end{equation} 
In addition, the leptonic Jarlskog is calculated to be $\text{J}_{\ell} = -0.0108$ and the neutrinoless double beta decay matrix element $\left<m_{\beta\beta}\right> = 0.00662$ eV.

\section{Conclusion}
\label{sec:conclude}
We modify the $SU(5)\times T^{\prime}$ GUT model proposed in~\cite{Chen:2007afa} with the additional singlet field $\eta^{\prime \prime}$, and a larger value for the lepton mixing angle $\theta_{13} \simeq 8^{\circ}-10^{\circ}$ can be accommodated. The analytical expressions of the lepton mixing angles and neutrino masses as well as various experimental observables including the Dirac CP phase $\delta$, Jarlskog invariant $J_{\ell}$  and the neutrinoless double beta decay matrix element $\left<m_{\beta\beta}\right>$ are derived. We present the numerical results in this letter and find large parameter space can satisfy all current experimental constraints. We also learn that it is easier to realize normal neutrino mass hierarchy than inverted neutrino mass hierarchy in this model. The illustrated numerical results of the model with the minimum $\chi^2$ fitting are consistent with our analytical expressions.

\section*{Acknowledgement}
We thank Michael Ratz for useful comments. The work of M-CC and A.M.W. was supported, in part, by the National Science Foundation under Grant No. PHY-0970173. The work of KTM was supported, in part, by the Department of Energy under Grant No. DE-FG02-04ER41290. JH is supported by the DOE Office of Science and the LANL LDRD program. A.M.W. would also like to acknowledge the hospitality of the TASI Summer School, where part of this work was completed.

\end{document}